# Phase-controlled Optical PT symmetry and asymmetric light diffraction in one- and two-dimensional optical lattices

Ali Akbar Naeimi[1], Elham Darabi[1], Ali Mortezapour[2*] Ghasem Naeimi[3]

[1]Department of physics, Science and Research Branch, Islamic Azad University, Tehran, Iran

[2]Department of Physics, University of Guilan, P. O. Box 41335–1914, Rasht, Iran

[3]Department of Physics, Qazvin Branch, Islamic Azad University, Qazvin, Iran

Corresponding author E-mail: mortezapour@guilan.ac.ir

**Abstract:**

We propose a novel scheme for asymmetric light diffraction of a weak probe field in a one-dimensional (1D) and two-dimensional (2D) lattice occupied with cold atoms. The atoms are driven into the double lambda-type configuration by a standing wave, two coupling laser fields and a probe. Our study suggests the proposed scheme is capable of forming an asymmetric diffraction as a result of inducing optical parity-time symmetry in the both 1D and 2D lattices. Moreover it is demonstrated that, the asymmetric pattern of diffraction can be dynamically manipulated by means of adjusting the relative phase. Furthermore it is revealed that in the case of 1D lattice (grating), Intensity variation of the coupling fields has a significant impact on the intensity of diffraction orders.

## 1- Introduction

Parity-Time (PT) symmetry is an active area of research in physics owing to its significance in the realm of science and technology. Bender and Boettcher [1] had pioneering role in introducing concept of the Parity-Time (PT) symmetry (1998) in quantum mechanics. A decade later, the reports indicated that optical systems can be considered as a candidate to provide a foundation for experimental investigation of *PT*-symmetric ideas. Moreover, It has been demonstrated that the necessary condition for realization of PT-symmetry in optical systems relies on satisfying $n(\vec{r}) = n^*(-\vec{r})$ term, which means the real and imaginary parts of complex refractive index respectively must be an even and odd function of $\vec{r}$. So far the experimental realization of the optical *PT*-symmetry in various structures such as lattices [2-5], waveguides [6, 7] and micro-cavities [8, 9] has been reported. Furthermore it has found potential applications in divers fields like as unidirectional propagation [10-13], lasing [14-18], perfect absorbers [19-21] and sensors [22]. In this context, a couple of interesting phenomena such as Bloch oscillations [23], electromagnetically induced transparency (EIT) [24], PT-symmetric Talbot effect [25], Giant Goos-Hänchen shift [26, 27] and optical solitons [28] have also been investigated in optical PT-symmetric structures.

In 1998, it was revealed that as a result of substituting a standing wave for the traveling wave of EIT, a diffraction grating could be formed in the atomic medium which yields the Fraunhofer diffraction of the weak probe field [29]. The phenomenon is known as electromagnetically induced grating (EIG). It is well-known that the formed gratings in ordinary media typically diffract light symmetrically [29-44] whereas in the PT-symmetric media it can have an asymmetric diffraction pattern since light undergoes spatially modulated refractive index [45]. In recent years, the realization of one-dimensional and two-dimensional asymmetric gratings in optical PT-symmetry structures has drawn the attention of many researchers [45-56]. Among the aforementioned structures, experimental feasibility of the optical lattices makes them superior to the other ones, hence a major portion of researches is devoted to them. Regarding the literature, in the all examined atomic lattices the atoms are driven in the either N-type or lambda-type configurations. However in our proposed scheme, we consider one-dimensional and two-dimensional lattices occupied with cold atoms which are driven in double lambda-type configurations with a closed-loop transition under the action of four applied fields. Here we have demonstrated that our novel scheme is capable of forming an asymmetric diffraction as a result of inducing optical parity-time symmetry. Moreover it is shown that the asymmetric pattern of diffraction can be dynamically manipulated by means of adjusting the relative phase of the applied fields.

The structure of the paper is as follows: In Sec. 2 we describe our model and provide an analytical solution for the density matrix equation. In Sec. 3 the details of one-dimensional grating and its results are presented. The results of two-dimensional grating are separately discussed in Sec. 4. Finally, Sec. 5 gives an outline of the study.

**2-Model and equations**

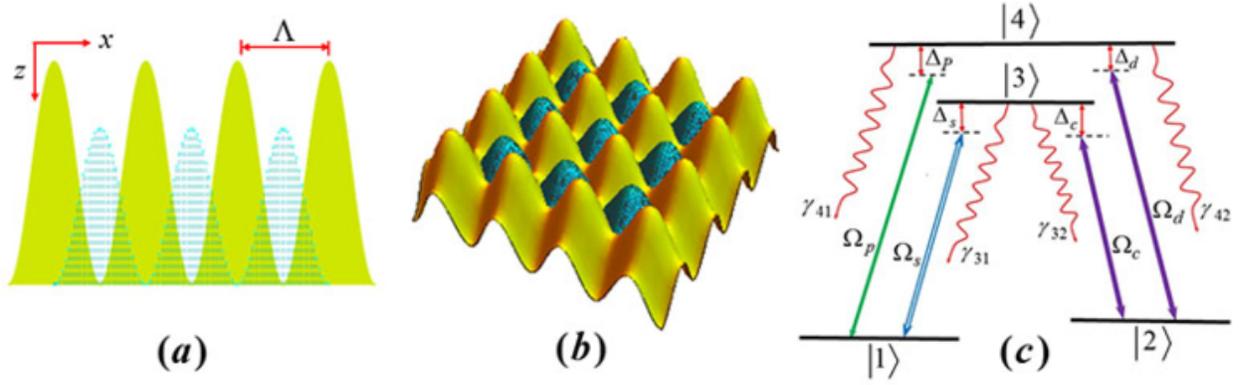

**Fig. 1.** (a) Sketch of 1D optical lattices with the period of $\Lambda$ along the *x* direction in which the blue area characterizes spatial distribution of the atomic density in dipole traps. (b) Sketch of 2D optical lattices. (c) Schematic diagram of the four-level double-lambda atomic system interacting with four applied fields in a closed-loop transition.

As shown in Fig. 1(a) and Fig.1 (b), we consider 1D and 2D optical lattices occupied with cold atoms at the bottom of each dipole trap. It is assumed each atom is driven into the four-level double-lambda configuration with a closed-loop transition by four coherent fields of frequencies (amplitudes and initial phase) $\omega_p$ ($E_p, \phi_p$), $\omega_c$ ($E_c, \phi_c$), $\omega_d$ ($E_d, \phi_d$) and $\omega_s$ ($E_s, \phi_s$). The atomic system is schematically depicted in Fig. 1(c). As can be seen, a probe field with Rabi frequency $\Omega_p = E_p \mu_{14}/2\hbar$ probes the transition $|1\rangle \leftrightarrow |4\rangle$, while the transitions $|2\rangle \leftrightarrow |3\rangle$ and $|2\rangle \leftrightarrow |4\rangle$ are driven by two strong control fields with Rabi frequencies $\Omega_c = E_c \mu_{23}/2\hbar$ and $\Omega_d = E_d \mu_{24}/2\hbar$ respectively. Moreover, a position-dependent coupling field with Rabi frequency $\Omega_s = E_s \mu_{13}/2\hbar$ is set to act on the transition $|1\rangle \leftrightarrow |3\rangle$. Here $\mu_{13}, \mu_{14}, \mu_{23}$ and $\mu_{24}$ are the corresponding electric-dipole matrix.

The Hamiltonian describing the dynamics of the system in dipole and rotating-wave approximation is given by:

$$\hat{H} = \sum_{j=1}^{4} \hbar \omega_j |j\rangle\langle j| - \{\hbar \Omega_s e^{-i(\omega_s t - \phi_s)}|3\rangle\langle 1| + \hbar \Omega_c e^{-i(\omega_c t - \phi_c)}|3\rangle\langle 2| \\ + \hbar \Omega_d e^{-i(\omega_d t - \phi_d)}|4\rangle\langle 2| + \hbar \Omega_p |4\rangle\langle 1| e^{-i(\omega_p t - \phi_p)} + H.c\} \tag{1}$$

The energy of the atomic level $|j\rangle$ is characterized by $\hbar \omega_j$. Switching to the interaction picture, the above Hamiltonian will be represented as follows (taking $\hbar = 1$):

$$V_I = (\Delta_c - \Delta_s)|2\rangle\langle 2| - \Delta_s|3\rangle\langle 3| + (\Delta_c - \Delta_s - \Delta_d)|4\rangle\langle 4| \\ - [\Omega_s(x)|3\rangle\langle 1| + \Omega_c|3\rangle\langle 2| + \Omega_d|4\rangle\langle 2| + \Omega_p|4\rangle\langle 1|e^{-i(\Delta t - \phi)} + H.c], \tag{2}$$

Here, $\Delta_s = \omega_s - \omega_{31}$, $\Delta_p = \omega_p - \omega_{41}$, $\Delta_c = \omega_c - \omega_{32}$, and $\Delta_d = \omega_d - \omega_{42}$, where $\omega_{ij}$ is the transition frequency between levels $|i\rangle$ and $|j\rangle$. We have also defined the so-called multiphoton resonance detuning $\Delta$ and relative phase $\phi$ as:

$$\Delta = (\Delta_p + \Delta_c) - (\Delta_s + \Delta_d), \tag{3}$$

$$\phi = (\phi_p + \phi_c) - (\phi_s + \phi_d). \tag{4}$$

The Liouville equations for the density matrix elements will be straightforwardly obtained by making use of Eq. (2):

$$\begin{aligned}
\dot{\rho}_{11} &= 2\gamma_{41}\rho_{44} + 2\gamma_{31}\rho_{33} + i\Omega_s^*\rho_{31} - i\Omega_s\rho_{13} + i\Omega_p^*\rho_{41} - i\Omega_p\rho_{14}, \\
\dot{\rho}_{22} &= 2\gamma_{32}\rho_{33} + 2\gamma_{42}\rho_{44} + i\Omega_c^*\rho_{32} - i\Omega_c\rho_{23} + i\Omega_d^*\rho_{42} - i\Omega_d\rho_{24}, \\
\dot{\rho}_{33} &= -(2\gamma_{31} + 2\gamma_{32})\rho_{33} + i\Omega_s\rho_{13} - i\Omega_s^*\rho_{31} + i\Omega_c\rho_{23} - i\Omega_c^*\rho_{32}, \\
\dot{\rho}_{12} &= i(\Delta_c - \Delta_s)\rho_{12} + i\Omega_s^*\rho_{32} - i\Omega_c\rho_{13} + i\Omega_p^*\rho_{42}e^{i(\Delta t - \phi)} - i\Omega_d\rho_{14}, \\
\dot{\rho}_{13} &= -(\Gamma_{31} + i\Delta_s)\rho_{13} + i\Omega_s^*(\rho_{33} - \rho_{11}) - i\Omega_c^*\rho_{12} + i\Omega_p^*e^{i(\Delta t - \phi)}\rho_{43}, \\
\dot{\rho}_{14} &= -[\Gamma_{41} - i(\Delta_c - \Delta_s - \Delta_d)]\rho_{14} + i\Omega_p^*e^{i(\Delta t - \phi)}(\rho_{44} - \rho_{11}) + i\Omega_s^*\rho_{34} - i\Omega_d^*\rho_{12}, \\
\dot{\rho}_{23} &= -[\Gamma_{32} + i\Delta_c]\rho_{23} + i\Omega_c^*(\rho_{33} - \rho_{22}) + i\Omega_d^*\rho_{43} - i\Omega_s^*\rho_{21}, \\
\dot{\rho}_{24} &= -[\Gamma_{42} + i\Delta_d]\rho_{24} + i\Omega_c(\rho_{44} - \rho_{22}) - i\Omega_p^*e^{i(\Delta t - \phi)}\rho_{21} + i\Omega_c^*\rho_{34}, \\
\dot{\rho}_{34} &= -[\Gamma_{43} + i(\Delta_d - \Delta_c)]\rho_{34} + i\Omega_s\rho_{14} + i\Omega_c\rho_{24} - i\Omega_p^*e^{i(\Delta t - \phi)}\rho_{31} - i\Omega_d^*\rho_{32}, \\
\rho_{11} &+ \rho_{22} + \rho_{33} + \rho_{44} = 1.
\end{aligned} \tag{5}$$

In the above density matrix equation, the phenomenological added overall dephasing rates $\Gamma_{ij}$ are given by $\Gamma_{31} = \gamma_{31} + \gamma_{32}$, $\Gamma_{32} = \gamma_{31} + \gamma_{32}$, $\Gamma_{41} = \gamma_{41} + \gamma_{42}$, $\Gamma_{42} = \gamma_{31} + \gamma_{41} + \gamma_{42}$ and $\Gamma_{43} = \gamma_{31} + \gamma_{32} + \gamma_{41} + \gamma_{42}$ where, as shown in Fig. 1(c), $\gamma_{ij}$ is the spontaneous decay rate from level $|i\rangle$ to the level $|j\rangle$. It is noteworthy that the multiphoton resonance condition in the above density matrix equation can be fulfilled by taking $\Delta = 0$.

In the particular case $\Delta_s = \Delta_p = \Delta_c = \Delta_d = 0$, $\gamma_{31} = \gamma_{32} = \gamma_{42} = \gamma_{41} = \gamma$ and with taking the weak probe and standing wave into account, the analytical steady-state solution of Eq. (5) yields the following equation for $\rho_{41}$:

$$\rho_{41} = -\frac{A}{2\gamma(B + C + D)} \tag{6}$$

where

$$A = ie^{2i\phi}\Omega_c\Omega_d^2(-\Omega_c^2\Omega_p + e^{i\phi}\Omega_c\Omega_d\Omega_s + 2\Omega_p\Omega_s^2),$$
$$B = -2\Omega_c^3\Omega_p^2 + 5e^{i\phi}\Omega_c^2\Omega_d\Omega_P\Omega_s,$$
$$C = e^{3i\phi}\Omega_c\Omega_d(\Omega_c^2\Omega_d + \Omega_d^3 + \Omega_c\Omega_p\Omega_s + 2\Omega_d\Omega_s^2),$$
$$D = -2e^{2i\phi}(\Omega_c^3\Omega_p^2 - 2\Omega_d^3\Omega_p\Omega_s + \Omega_c\Omega_d^2\Omega_s^2).$$

within range of the weak probe field, the local steady-state probe susceptibility of the medium can be acquired as

$$\chi_p = \frac{|\vec{\mu}_{41}|^2 N_j}{\hbar\varepsilon_0\Omega_p}\rho_{41}, \tag{7}$$

Where $\varepsilon_0$ indicates the dielectric constant in vacuum and $N_j$ denotes the position-dependent atomic density at each trap.

Since the susceptibility is a complex parameter, one can consider it as $\chi_p = \text{Re}[\chi_p] + i\,\text{Im}[\chi_p]$ in which the real ($\text{Re}[\chi_p]$) and imaginary ($\text{Im}[\chi_p]$) parts describe the dispersion and absorption properties of the probe field, respectively. Note that, $\text{Im}[\chi_p] < 0$ ($\text{Im}[\chi_p] > 0$) indicates gain (loss). On the other hand, To attain a PT symmetry in photonic systems, the complex refractive index is necessary to meet the condition $n(\vec{r}) = n^*(-\vec{r})$ which implies such a system is constituted by spatially modulating balanced gain and loss. It is noteworthy that the refractive index is associated with the probe susceptibility via $n = \sqrt{1+\chi_p} \approx 1 + \chi_p/2$. Thus regarding the aforementioned facts, one shall deduce the real part of the refractive index ($\text{Re}[\chi_p]$) has a symmetric pattern, while the imaginary part ($\text{Im}[\chi_p]$) is antisymmetric.

**3- 1D atomic grating:**

In order to achieve 1D atomic grating, the optical lattices are assumed to be arranged in 1D with the period of $\Lambda$ along the $x$ direction as shown in Fig. 1(a). Meanwhile, the atomic density in the jth dipole trap has a Gaussian distribution as

$$N_j(x) = N_0 e^{-(x-x_j)^2/\sigma^2}, x \in (x_j - \Lambda/2, x_j + \Lambda/2). \tag{8}$$

Where $N_0$ is the (average) peak density, $\sigma$ denotes the half-widths of the Gaussian distribution and $x_j$ indicates the trap center. Furthermore, the position-dependent coupling field $\Omega_s$ is considered to be a superposition of a traveling-wave (TW) field and a standing-wave (SW) field along $x$ direction which can be written as $\Omega_s(x) = \Omega_{s_0} + \delta\Omega_s \sin[2\pi x/\Lambda]$.

Now, one can rewrite the local steady-state probe susceptibility of the medium (Eq. (6)) as

$$\chi_p = \left(\frac{|\vec{\mu}_{41}|^2 N_0}{\hbar\varepsilon_0\gamma_{41}}\right)\chi'_p(x), \tag{9}$$

In which $\chi'_p(x) = \left(\frac{\rho_{41}\gamma_{41}}{\Omega_p}\right)e^{-(x^2-x_j^2)/\sigma^2}$ characterizes the normalized local susceptibility.

To obtain the self-consistent equation for propagation of the probe field through the medium, we restrict ourselves to the approximation of the slowly varying envelope of the probe field [57]. This enables us to consider Maxwell's equation [58] for the probe field with the wavelength $\lambda_p$ in the following form,

$$\frac{\partial E_p}{\partial z} = i\frac{\pi}{\lambda_p}\chi_p E_p, \tag{10}$$

By solving Eq. (10), the normalized transmission function $T(x) = E_p(z=L)/E_p(z=0)$ for the interaction length $L$ of the medium (grating thickness) can be accomplished as:

$$T(x) = e^{-\mathrm{Im}[\chi'_p(x)]L/\xi} e^{i\mathrm{Re}[\chi'_p(x)]L/\xi}. \tag{11}$$

Where $\xi = \hbar\varepsilon_0\gamma_{41}/N_0|\mu_{14}|^2 k_p$. Here the terms $e^{-\mathrm{Im}[\chi'_p(x)]L/\xi}$ and $e^{i\mathrm{Re}[\chi'_p(x)]L/\xi}$ are amplitude and phase of the transmission function $T(x)$. The probe field is assumed to be a plane wave, thus intensity distribution in the far field (Fraunhofer diffraction) will be [1]:

$$I_p(\theta_x) = |E(\theta_x)|^2 \left[\frac{\sin^2(\pi MR\sin\theta_x)}{M^2\sin^2(\pi R\sin\theta_x)}\right]. \tag{12}$$

Where $R = \Lambda/\lambda_p$, $M$ denotes the ratio between the probe beam width and the grating period. Moreover $\theta_x$ represents the diffraction angle along the $z$-direction in the $x-z$ plane, whereas the quantity $E(\theta_x)$ stands for the far-field diffraction amplitude stemming from one single lattice, which reads

$$E(\theta_x) = \int_{-\Lambda/2}^{\Lambda/2} T(x)\exp[-i2\pi Rx\sin\theta_x]dx. \tag{13}$$

Let us start the discussion with a study of the local normalized dispersion and absorption of the probe field based on Eq. (6). Fig. 2 displays $\mathrm{Im}[\chi'_p]$ and $\mathrm{Re}[\chi'_p]$ versus the scaled coordinate $(x-x_j)/\Lambda$ for different relative phase of the applied beams. Regarding the antisymmetric profile of $\text{Im}[\chi'_p]$ for $\phi = \pi/2$ and $\phi = 3\pi/2$, one can deduce that optical PT-symmetry is appreciably induced to the system, while the diagrams of $\phi = 0$ and $\phi = \pi$ reveal that the condition of optical PT-symmetry is not satisfied. Moreover, it is worth noticing that in spite of the identical pattern of $\text{Im}[\chi'_p]$ for $\phi = \pi/2$ and $\phi = 3\pi/2$, the patterns of $\text{Re}[\chi'_p]$ are totally inverse. The interpretation of such behavior is presented in Fig. 3.

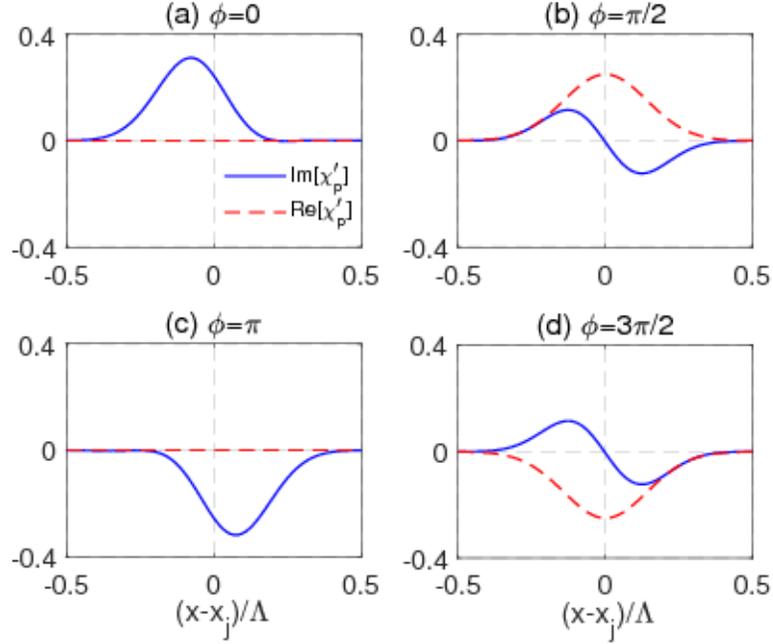

**Fig. 2.**

In order to emphasize the importance of the relative phase on the diffraction pattern of the probe, we plot $I_p(\theta_x)$ versus $\sin\theta_x$ in Fig. 3. It is evident that for $\phi = 0$, due to the negligible value of $\text{Re}[\chi'_p]$ only a near-amplitude absorption grating is formed. However, for $\phi = \pi/2$ and $\phi = 3\pi/2$, asymmetric gain phase-grating are observed in which mirror inversion of the diffraction patterns is due to the opposite signs of $\text{Re}[\chi'_p]$ (dispersion) for these phases.

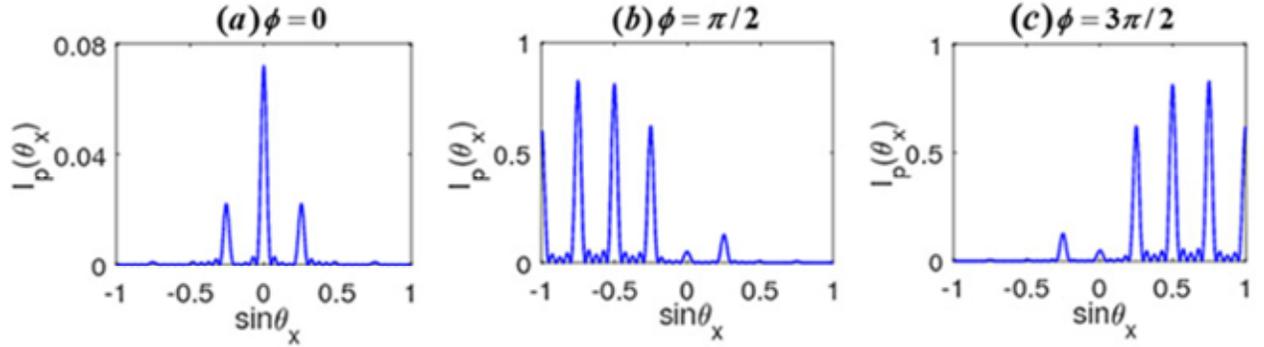

**Fig. 3.**

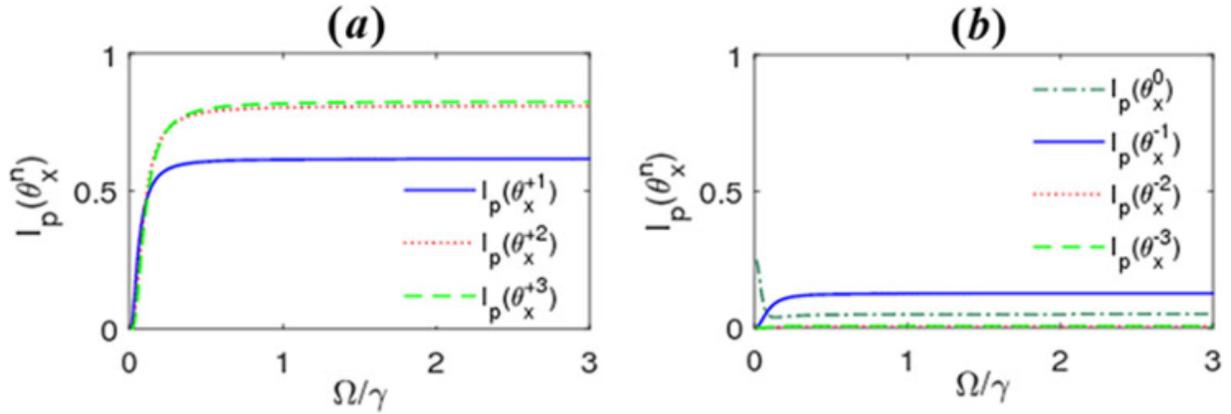

**Fig. 4.**

Now, we aim to assess the impact of the coupling fields on the intensity of different (zeroth, negative and positive) diffraction orders by means of numerically solving of the Eq. (4). In Fig. 4, it is assumed $\Omega_c = \Omega_d = \Omega$ and the other parameters are the same as those used in Fig. 3(c). As it is obvious by applying the coupling fields and increasing the intensities ($\Omega$), all the positive diffraction orders and $I_p(\theta_x^{-1})$ show a similar behavior. So that firstly they have an increasing trend then they take a constant value. Although compared to the appreciable increase of the positive diffraction orders, the increasing trend is small for $I_p(\theta_x^{-1})$. In contrast, $I_p(\theta_x^0)$ starts with a decreasing trend then takes a minor constant value. Meanwhile, $I_p(\theta_x^{-2})$ and $I_p(\theta_x^{-3})$ show abnormal behavior. As it is manifest these diffraction orders behave indifferently to the coupling field and its increase. Note that the asymmetric pattern of diffraction is achievable for $\Omega \geq 0.1\gamma$.

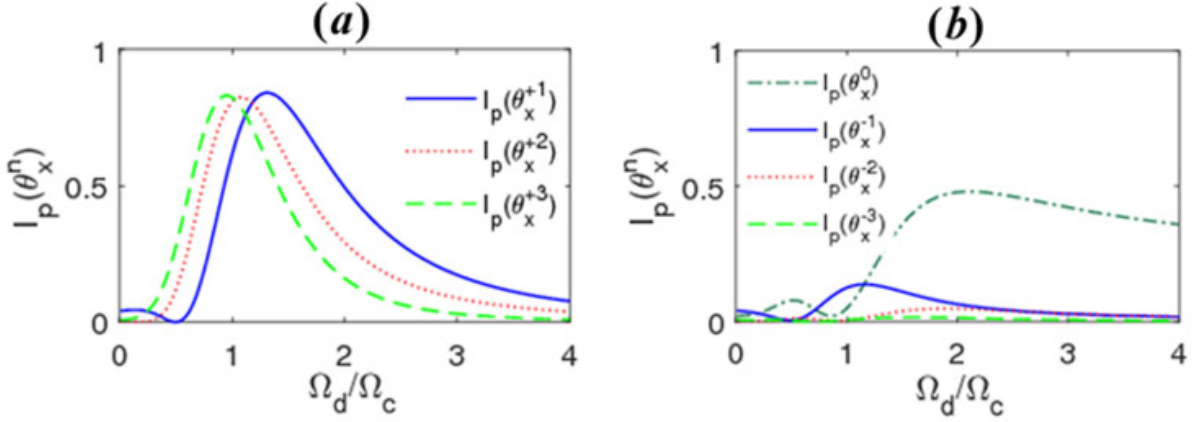

**Fig. 5**

In Fig. 5 the effect of the relative Rabi frequency ($\Omega_d / \Omega_c$) of the coupling fields on the intensity of (zeroth, negative and positive) diffraction orders is examined. This figure discloses that the intensity of the diffraction orders which had been appeared in Fig. 3(c), can be conveniently manipulated by varying $\Omega_d / \Omega_c$ which leads to accomplishing an optimum condition. Namely, as the positive-orders become significantly amplified, the other diffraction orders enormously attenuate. Moreover, the diagram reveals that the large $\Omega_d / \Omega_c$ is unfavorable because destroys the asymmetric pattern of the diffraction and makes the lattice function as an amplitude grating.

**4- 2D atomic grating:**

In this section, we extend the model to the case of 2D optical lattices in which cold driven atoms are located into the dipole traps with density distribution of $N_j(x) = N_0 e^{-[(x-x_j)^2/\sigma_x^2 + (y-y_j)^2/\sigma_y^2]}$ at each trap (see Fig.1 (b)). Here, $\sigma_x$ ($\sigma_y$) is the half-width of the Gaussian profile along the $x$ ($y$)-direction. Moreover, the transition $|1\rangle \leftrightarrow |3\rangle$ is required to be derived by the position-dependent coupling field with Rabi frequency $\Omega_s(x,y) = \Omega_{s_0} + \delta\Omega_s(\sin[2\pi x/\Lambda_x] + \sin[2\pi y/\Lambda_y])$. Note that in this case, the position-dependent coupling field is the superposition of a traveling-wave (TW) field and two standing-wave (SW) fields along $x$ and $y$ directions.

In a similar way of 1D grating, the normalized local susceptibility of the system can be acquired as:

$$\chi'_p(x,y) = \left(\frac{\rho_{41}\gamma_{41}}{\Omega_p}\right) e^{-[(x^2-x_j^2)/\sigma_x^2 + (y-y_j)^2/\sigma_y^2]} \tag{14}$$

Then, the transmission function of the probe field can be written as $T(x,y) = e^{-\text{Im}[\chi'_p(x,y)]L/\xi} e^{i\text{Re}[\chi'_p(x,y)]L/\xi}$. Making use of the Fourier transform of the transmission function, the Fraunhofer 2D diffraction intensity of the probe field, propagating perpendicular to the atomic lattice, can be written as,

$$I(\theta_x,\theta_y) = |E(\theta_x,\theta_y)|^2 \left[\frac{\sin^2(\pi M_x R_x \sin\theta_x)}{M_x^2 \sin^2(\pi R_x \sin\theta_x)} \frac{\sin^2(\pi M_y R_y \sin\theta_y)}{M_y^2 \sin^2(\pi R_y \sin\theta_y)}\right]. \tag{15}$$

Where

$$E(\theta_x,\theta_y) = \int_{-\Lambda_x/2}^{\Lambda_x/2} dx \int_{-\Lambda_y/2}^{-\Lambda_y/2} T(x,y) e^{-i2\pi(R_x x \sin\theta_x + R_y y \sin\theta_y)} dy, \tag{16}$$

$R_x = \Lambda_x/\lambda_p$, $R_y = \Lambda_y/\lambda_p$, $\theta_y(\theta_x)$ being the diffraction angle with respect to the z-direction in the $y(x)-z$ plane and $M_{y(x)}$ stands for the number of spatial periods of the 2D atomic grating along the $y(x)$-direction. However, in this study, the optical lattice is assumed to be square ($\Lambda_x = \Lambda_y = \Lambda$, $M_x = M_y = M$).

We first analyze the imaginary (gain or absorption) and the real (dispersion) parts of the normalized susceptibility $\chi'_p$ as functions of the scaled coordinates $(x-x_j)/\Lambda$ and $(y-y_j)/\Lambda$ for different relative phase of the applied fields in Fig. 6. As the contour of $\text{Im}[\chi'_p]$ for $\phi = 0$ displays, there are absorption and gain in some areas which do not conform the symmetric or antisymmetric pattern and the dispersion is zero. However $\text{Im}[\chi'_p]$ for $\phi = \pi/2$ and $\phi = 3\pi/2$ demonstrates that there is an antisymmetric relation between absorption and gain whereas, $\text{Re}[\chi'_p]$ for both $\phi = \pi/2$ and $\phi = 3\pi/2$ fits the symmetric pattern with taking mirror inversion into account. These results obviously suggest that 2D PT symmetry can be realizable in the 2D lattice by well-adjusting the relative phase of the applied fields.

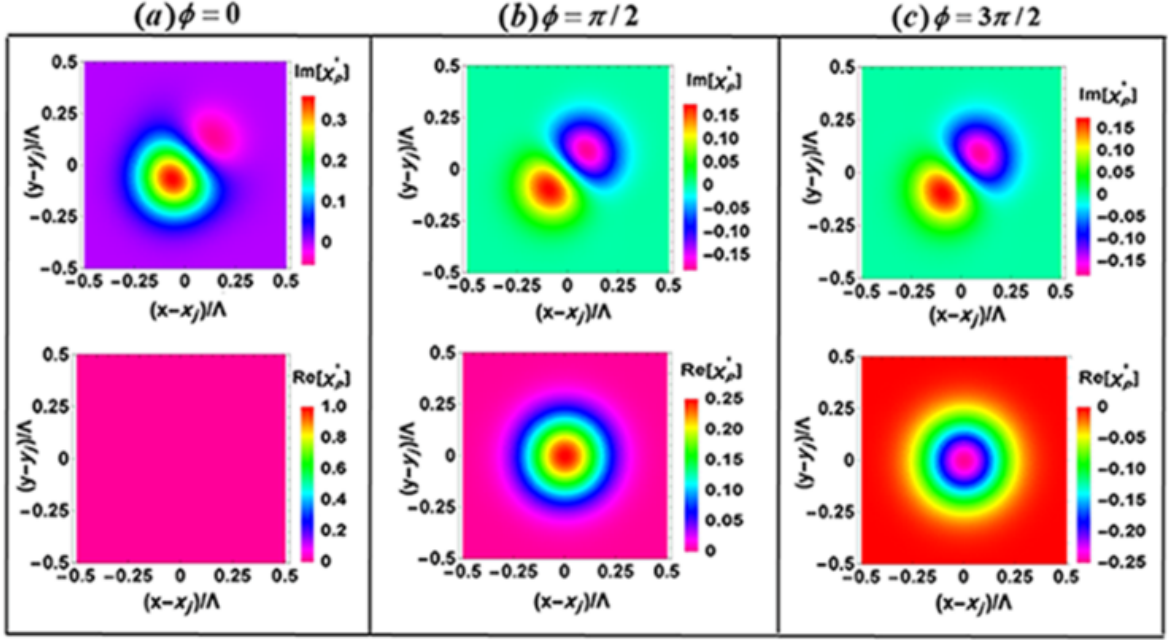

**Fig. 6.**

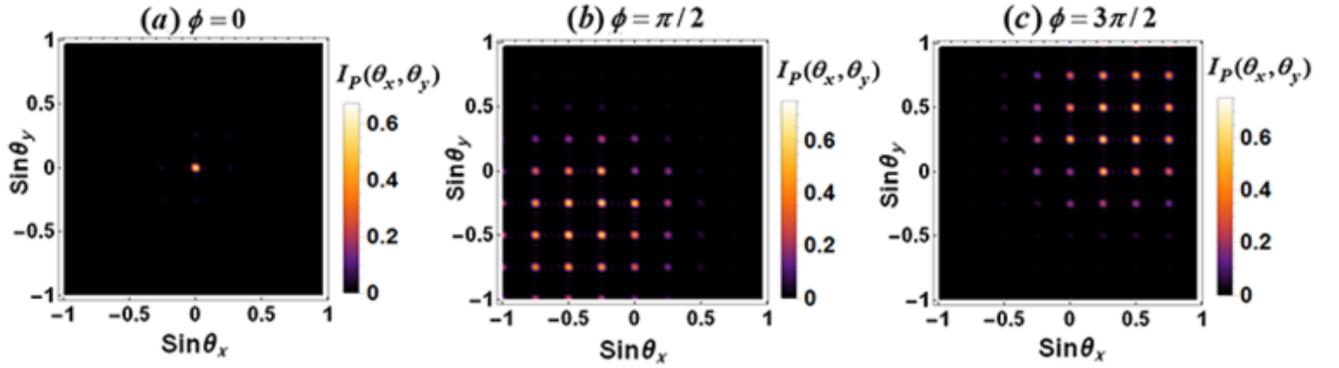

**Fig. 7**

In Fig. 7, three-dimensional plots of the Fraunhofer diffraction patterns ($I(\theta_x,\theta_y)$) corresponding to $\sin\theta_x$ and $\sin\theta_y$ is depicted for (a) $\phi=0$, (b) $\phi=\pi/2$ and (c) $\phi=3\pi/2$. As can be seen a highly centralized structure of the diffraction for the case of $\phi=0$, implies formation of an absorption amplitude grating. However as the Fig. 7(b) illustrates for $\phi=\pi/2$, the probe field diffraction has been decentralized in $-1\leq\sin\theta_x,\sin\theta_y\leq 0$ region with an approximately even distribution of intensity. All the descriptions for $\phi = \pi/2$ hold true for $\phi = 3\pi/2$, with only one difference that the distribution region is shifted to $0 \leq \sin\theta_x, \sin\theta_y \leq 1$. The difference is originated from the opposite sign of the dispersion.

**4- Conclusion:**

In summary, the asymmetric Fraunhofer diffraction of a weak probe field passing through a one-dimensional (1D) and two-dimensional (2D) lattices occupied with cold atoms have been investigated. It is assumed the atoms are driven into the closed loop four-level double-lambda configuration by the probe, a standing wave and two coupling fields. Our study suggests that in both 1D and 2D lattices, relative phase of the applied fields can be used as a controlling parameter to manipulate the optical properties of the medium and accomplish an optimum condition. It is revealed that as a result of adjusting the relative phase, optical parity-time symmetry can be induced to the system which gives rise to the asymmetry of diffraction. It is worth mentioning that the asymmetric pattern of diffraction is accompanied by gain which certifies the efficiency of the proposed scheme. Furthermore, in the case of 1D lattice (grating) it has been disclosed that the intensity of different diffraction orders in the asymmetric pattern is highly affected by varying the intensities of the coupling fields. Inducing a PT symmetry in a double-lambda configuration interacting with four fields is a fresh idea which has not previously been reported. The convenient manipulation of the asymmetric diffraction pattern by means of the relative phase demonstrates superiority of the proposed scheme.

**Disclosures:**

The authors declare no conflicts of interest related to this paper.

**Figures caption**

**Fig. 1.** (a) Sketch of 1D optical lattices with the period of $\Lambda$ along the $x$ direction in which the blue area characterizes spatial distribution of the atomic density in dipole traps. (b) Sketch of 2D optical lattices. (c) Schematic diagram of the four-level double-lambda atomic system interacting with four applied fields in a closed-loop transition.

**Fig. 2.** Plots of normalized absorption $\text{Im}[\chi'_p]$ (blue line) and dispersion $\text{Re}[\chi'_p]$ (red-dotted line) as functions the scaled coordinate $(x-x_j)/\Lambda$ for (a) $\phi = 0$, (b) $\phi = \pi/2$ and (c) $\phi = 3\pi/2$. Other specific parameters are $\Omega_p = 0.05\gamma$, $\Omega_{s_0} = 0.001\gamma$, $\delta\Omega_s(x) = 0.05\gamma$, $\Omega_c = \Omega_d = 2\gamma$, $\sigma = 0.2\Lambda$, $\Delta_s = \Delta_p = \Delta_c = \Delta_d = 0$, $\gamma_{31} = \gamma_{32} = \gamma_{41} = \gamma_{42} = \gamma$.

**Fig. 3.** 1D Diffraction pattern $I_p(\theta_x)$ versus $\sin\theta_x$ for (a) $\phi = 0$, (b) $\phi = \pi/2$ and (c) $\phi = 3\pi/2$. Other specific parameters are $\Omega_p = 0.05\gamma$, $\Omega_{s_0} = 0.001\gamma$, $\delta\Omega_s(x) = 0.05\gamma$, $\Omega_c = \Omega_d = 2\gamma$, $L = 20\xi$, $\sigma = 0.2\Lambda$, $M = 5$, $\Lambda/\lambda_p = 4$, $\Delta_s = \Delta_p = \Delta_c = \Delta_d = 0$ and $\gamma_{31} = \gamma_{32} = \gamma_{41} = \gamma_{42} = \gamma$.

**Fig. 4.** The intensity of different (zeroth, negative and positive) diffraction orders as functions of $\Omega_c = \Omega_d = \Omega$. Values of the other parameters are taken as follows: $\Omega_p = 0.05\gamma$, $\Omega_{s_0} = 0.001\gamma$, $\delta\Omega_s(x) = 0.05\gamma$, $\phi = 3\pi/2$, $L = 20\xi$, $\sigma = 0.2\Lambda$, $M = 5$, $\Lambda/\lambda_p = 4$, $\Delta_s = \Delta_p = \Delta_c = \Delta_d = 0$ and $\gamma_{31} = \gamma_{32} = \gamma_{41} = \gamma_{42} = \gamma$.

**Fig. 5.** The intensity of different (zeroth, negative and positive) diffraction orders as functions of $\Omega_d/\Omega_c$ with $\Omega_c = 2\gamma$. Other parameters are the same as those in Fig. 4.

**Fig. 6.** Plots of normalized absorption $\text{Im}[\chi'_p]$ and dispersion $\text{Re}[\chi'_p]$ as functions the scaled coordinates $(x-x_j)/\Lambda$ and $(y-y_j)/\Lambda$ for (a) $\phi = 0$, (b) $\phi = \pi/2$ and (c) $\phi = 3\pi/2$. Other specific parameters are $\Omega_p = 0.05\gamma$, $\Omega_{s_0} = 0.001\gamma$, $\delta\Omega_s(x) = 0.05\gamma$, $\Omega_c = \Omega_d = 2\gamma$, $\sigma = 0.2\Lambda$, $\Delta_s = \Delta_p = \Delta_c = \Delta_d = 0$, $\gamma_{31} = \gamma_{32} = \gamma_{41} = \gamma_{42} = \gamma$.

**Fig. 7.** 2D Diffraction pattern $I_p(\theta_x, \theta_y)$ as functions $\sin\theta_x$ and $\sin\theta_y$ for (a) $\phi = 0$, (b) $\phi = \pi/2$ and (c) $\phi = 3\pi/2$. Other specific parameters are $\Omega_p = 0.05\gamma$, $\Omega_{s_0} = 0.001\gamma$, $\delta\Omega_s(x) = 0.05\gamma$, $\Omega_c = \Omega_d = 2\gamma$, $L = 20\xi$, $\sigma = 0.2\Lambda$, $M = 5$, $\Lambda/\lambda_p = 4$, $\Delta_s = \Delta_p = \Delta_c = \Delta_d = 0$, $\gamma_{31} = \gamma_{32} = \gamma_{41} = \gamma_{42} = \gamma$.